\documentclass[prb,twocolumn,floats]{revtex4}% Physical Review B
\usepackage{graphicx,epsfig}% Include figure files

\begin{document}

%\draft
%\preprint{}

\title{Resistance scaling for Composite Fermions in the presence of a density gradient}

\author{W. Pan}
\affiliation{Sandia National Laboratories, Albuquerque, New Mexico 87185}

\author{H.L. Stormer}
\affiliation{Department of Physics and Department of Applied Physics, Columbia University, New York, NY 10027, and\\
Bell Labs, Lucent Technologies, Murray Hill, NJ 07974}

\author{D.C. Tsui}
\affiliation{Department of Electrical Engineering, Princeton University, Princeton, NJ 08544}

\author{L.N. Pfeiffer, K.W. Baldwin, and K.W. West}
\affiliation{Bell Labs, Lucent Technologies, Murray Hill, New Jersey 07954}
%
%\vskip5pc

\date{\today}

\begin{abstract}

The magnetoresistance, $R_{xx}$, at even-denominator fractional fillings, of an ultra high quality two-dimensional electron
system at $T \sim 35$ mK is observed to be strictly linear in magnetic field, $B$. While at 35 mK $R_{xx}$ is dominated by
the integer and fractional quantum Hall states, at $T \cong 1.2$ K an almost perfect linear relationship between $R_{xx}$ and
$B$ emerges over the whole magnetic field range except for spikes at the integer quantum Hall states. This linear $R_{xx}$
cannot be understood within the Composite Fermion model, but can be explained through the existence of a density gradient in
our sample.

\end{abstract}


\maketitle

The Composite Fermion (CF) model \cite{jain89,lopez,hlr} has been very successful in explaining the Fermi liquid like
behavior at even denominator Landau level fillings in a two dimensional electron system (2DES). Within this model, a CF is
formed by attaching 2$\phi$ flux quanta to each electron at filling factor $\nu=1/2\phi$, where $\phi$ is positive integer.
Due to this flux attachment, the CFs see a zero effective $B$ field. Furthermore, since the strong electron-electron
interaction is effectively removed by forming the CFs, the resulting CF-CF interaction is very weak. Consequently, at low
temperatures, their transport properties can be well described by Fermi liquid theory.

Over the years, the CF model has been tested and verified in many types of experiments \cite{jain00}. However, a few
unresolved issues remain. Among them is the resistivity of CFs. Unlike ordinary electrons, whose main scattering mechanism is
Coulomb scattering from residual impurities within the samples, it is believed that for the CFs the main scattering mechanism
is due to the gauge field fluctuations introduced by the same impurities \cite{hlr}. Therefore, the resistivity of CFs is
given by the formula; $\rho^{CF}_{xx} \approx (n_{imp}/n) \times (\pi\phi^2/k_fd_se^2)$, where $n_{imp}$ is the density of
residual impurities, $n$ is the 2DES density, $k_f$ is the CF Fermi wavevector, and $d_s$ is the spacing of the impurities
from the 2DES. However, it has long been noticed \cite{hlr} that the experimental values at $\nu = 1/2$ are always smaller
than the theoretically predicted ones, typically by a factor of 3 and more. Although it was speculated \cite{halperin} that
this discrepancy might be related to the specifics of the density inhomogeneity the issue was never resolved.

Another discrepancy between theory and experiment in CF transport is found in the scaling of the resistivity for CFs of the
same flavor (or the same $\phi$), and/or of different flavors. According to the above equation, one would expect $R_{\nu=1/2}
= 3^{1/2} \times R_{\nu=3/2}$, and $R_{\nu=1/4} = 4 \times R_{\nu=1/2}$. So far, this CF resistivity scaling, especially the
one between $\nu = 1/2$ and 1/4, could not be clearly tested. The primary reason is that even in very high mobility samples
the 2DES often becomes insulating beyond $\nu =1/3$ \cite{jiang}. Consequently, the resistivity at $\nu = 1/4$ becomes very
large and a comparison becomes meaningless.

Recently many high order FQHE states, e.g., the $\nu =4/11$, 10/21 and 10/19 states around $\nu = 1/2$ and the $\nu =6/23$
and 6/25 states around $\nu = 1/4$, were observed \cite{pan03}. Such ultra high quality specimen allows for a reliable and in
depth investigation of the resistivity of CFs at the even-denominator fillings, i.e., $\nu =1/4$, 1/2, 3/4, and 3/2. As it
turns out, the resistivity at even-denominator filing factors is found to be linear in $B$ field, which is at variance from
standard CF transport theory. This linear magnetoresistance (MR) becomes very pronounced at a high temperature of $T \cong
1.2$~K, where only a few sharp spikes from the integer quantum Hall states disrupt an otherwise strictly linear relationship
between $R_{xx}$ and $B$. Such a linear MR is not consistent with the resistivity scaling from the CF model. However, all
such features can be understood assuming a small electron density gradient within the 2DES.

The sample consists of a symmetrically doped quantum well of width 500 \AA. The setback distance of the modulation doping is
$d_s$ = 2200 \AA. An electron density of $n \simeq 1 \times 10^{11}$ cm$^{-2}$ and a mobility $\mu \simeq 1 \times 10^7$
cm$^2$/Vs were achieved after illumination of the sample at low temperatures by a red light-emitting diode (LED). A
self-consistent calculation shows that at this density only one electrical subband is occupied. Conventional low-frequency
($\sim$ 7Hz) lock-in amplifier techniques were employed to measure the diagonal magnetoresistance $R_{xx}$ and Hall
resistance $R_{xy}$.

In Figure 1a, we plot the $R_{xx}$ data at $T \sim 35$ mK. This data was shown earlier in Ref. [7] in the context of the
discovery of new FQHE states. Here, we focus on the extended straight sections around $\nu =1/4$ and $\nu =1/2$. In most
previous experiments the data beyond $\nu =1/3$ experienced a considerable increase in resistivity, often tending towards
infinity as $T$ towards zero. Such divergent behavior is closely correlated with sample quality as measured by mobility and
is generally attributed to magnetic field induced localization, which is furthered by increased disorder. The absence of such
a rising background in our data and the lack of a temperature dependence in the $\nu =1/4$ regime attests to the ultra-high
quality of our sample. It renders this sample an excellent candidate for a study of CF transport behavior. In Fig. 1b, the
resistance at the even-denominator fillings $\nu =3/2$, 3/4, 1/2, and 1/4 from three different cool-downs are plotted versus
$B$ field. A linear dependence on $B$ field is clearly observed.

\begin{figure} [h]
\centerline{\epsfig{file=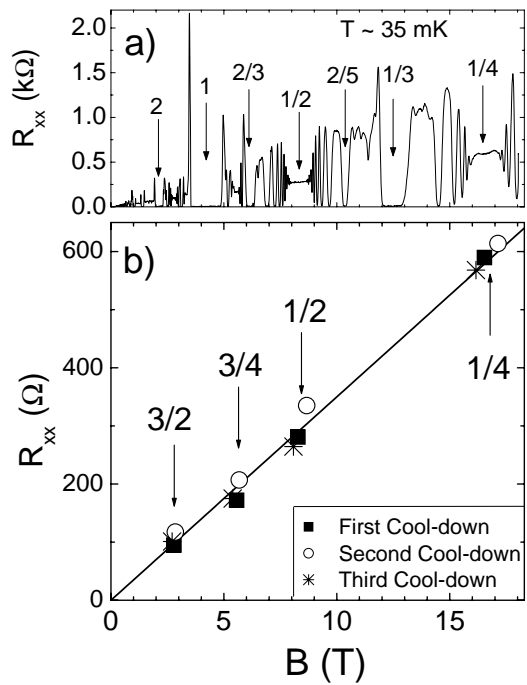,width=8.5cm}} \caption{(a) $R_{xx}$ trace at $T \sim 35$~mK. A detailed discussion on this
$R_{xx}$ data can be found in Ref. [7]. (b) $R_{xx}$ at the even-denominator fillings $\nu$ = 3/2, 3/4, 1/2, and 1/4 at $T
\sim 35$ mK in three different cool-downs. The 2D electron density depends on the LED illumination condition and varies a
little bit from one cool-down to another. The line is a guide to eye.}
\end{figure}

This linear $B$ field dependence becomes more extended at higher temperatures as seen in Fig. 2, where we show $R_{xx}$ and
$R_{xy}$ at $T \cong 1.2K$. At such high temperature, $R_{xy}$ behaves practically classically and is linear in $B$ field.
What is surprising is that $R_{xx}$ also shows a linear $B$ dependence over the whole $B$ field range, except at those
positions where the integer quantum Hall states start to form.

\begin{figure} [h]
\centerline{\epsfig{file=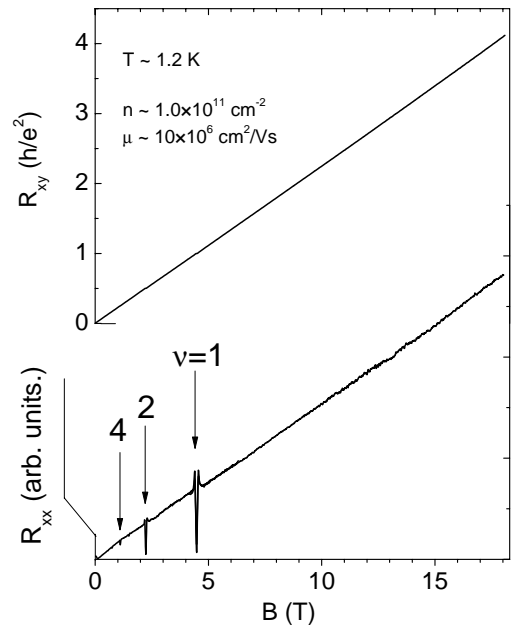,width=8.5cm}} \caption{Diagonal resistance $R_{xx}$ and Hall resistance $R_{xy}$ at $T \sim
1.2$ K. IQHE states are forming at the Landau level fillings $\nu$ = 1, 2, and 4}
\end{figure}

This linear MR is puzzling. First, it is inconsistent with the semiclassical theory of Lifshitz-Azbel-Kaganov
\cite{lifshitz}, which states that in a Fermi liquid with a close Fermi surface, MR should saturate when $\omega_c \tau =\mu
B \gg 1$, where $\omega_c = \hbar B/m^*$ is the cyclotron frequency, $\tau$ is the scattering time, and $m^*$ is the electron
effective mass. Therefore, in our specimen with $\mu \sim 1 \times 10^7$ cm$^2$/Vs, $R_{xx}$ should be constant beyond $B
\sim 10^{-3}$ T. Second, the linear MR cannot be understood within the CF model \cite{jain89,lopez,hlr}, which requires
$\rho^{CF}_{xx} \approx (n_{imp}/n) \times (\pi \phi^2/k_fd_se^2)$. Consequently, one would expect that $R_{1/4} = 4 \times
R_{1/2}$ and $R_{1/2} = 3^{1/2} \times R_{3/2}$. The experimental data, however, indicate $R_{1/4} = 2 \times R_{1/2}$, and
$R_{1/2} = 3 \times R_{3/2}$, which is clearly different. Linear MR in a 2DES has been observed before
\cite{rotger,hirai,stormer}. In fact, in one publication \cite{hirai}, the authors suggested that it might be related to a
density inhomogeneity.

To understand the physical origin of the linear MR, we draw from our recent publication \cite{pan05} on the empirical
resistivity rule \cite{simon}. There it was shown that $R_{xx}$ data in the first and second Landau levels turned out to be
merely a reflection of $R_{xy}$. This relationship was caused by an unintentional electron density gradient, $\Delta n$, in
the sample and expressed by an earlier, empirical resistivity rule

\begin{equation}
R_{xx} = R_{xy}(n) - R_{xy}(n + \Delta n) = c \times B \times dR_{xy}/dB
\end{equation}

where the constant is now determined to be $c=\Delta n/n$. Following this recent insight we examine our data in this light.
Obviously, an $R_{xy} \propto B$ leads to a linear $R_{xx}$, suggesting that its origin is again an unintentional electron
density gradient in our sample. Moreover, if we calculate $R_{xx}$ directly from the $R_{xy}$ data according to Eq.(1) we
reach practically perfect agreement with our $R_{xx}$ data (as shown in Fig.3), assuming a relative density gradient of $c =
\Delta n/n=0.5$\%. This is very strong evidence that the linear MR is a result of $R_{xy}$ via the resistivity rule, which
has been traced back to a density gradient.

\begin{figure} [h]
\centerline{\epsfig{file=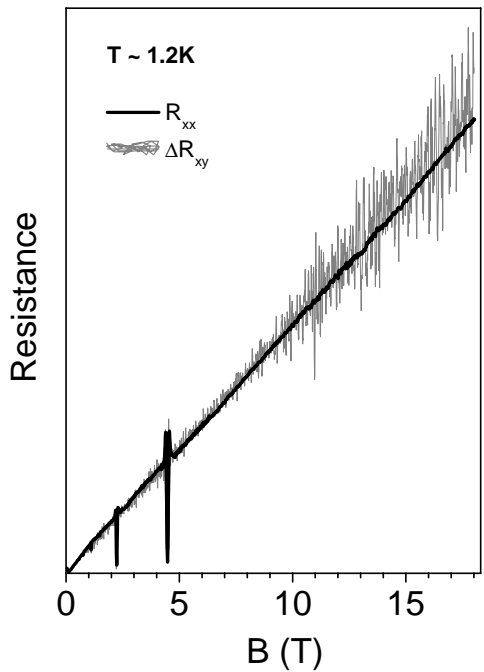,width=8.5cm}} \caption{Comparison of $R_{xx}$ (black trace) and $\Delta R_{xy}$ (noisy gray
trace).}
\end{figure}

It is remarkable that three apparently unrelated transport features, the linear MR, $R_{xx}$ quantization \cite{pan05}  and
the resistivity rule \cite{simon}, can all be explained by the empirical density gradient model, and that the explanation
holds well over an extremely wide temperature range, from 6~mK to 1.2~K and may well hold beyond this range. This simple,
classical explanation of $R_{xx}$ data in terms of $R_{xy}$ raises the question as to its connection to the intrinsic
resistivity $\rho_{xx}$. Recent theoretical work is addressing this relationship \cite{cooper}.

As to the resistivity at even denominator fillings and their theoretical relationships it is no longer surprising that they
are not borne out in experiment. $R_{xx}$ is the result of a density gradient and really only a reflection of $R_{xy}$.
Therefore, typical $R_{xx}$ data say little about $\rho_{xx}$ and hence little about the scattering behavior of the CFs. The
extraction of reliable $\rho_{xx}$ values from $R_{xx}$ data on 2DESs will require either the reduction or elimination of
residual electron density gradients or the application of appropriate correction formulae, based on realistic models for the
density distribution as they are presently being developed \cite{cooper}.

Before concluding we would like to put our data and their interpretation in the context of a wider scope: Recently a
semiclassical approach has been proposed to explain the linear magnetoresistance in 2DES \cite{polyakov}. Within this model,
a linear MR arises from a competition between the long-range and the short-range disorder potentials. This model was tested
in a 2DES with an antidot array, to create the short-range disorder potential \cite{renard}, and provided general support.
However, it appears unlikely that this model would also apply to our setting, since our sample is unpatterned. Moreover, in
antidots \cite{renard}, the linear MR was strong only at high temperatures, whereas, in our sample, the linear MR was
observed from 35~mK to our maximum temperature of 1.2~K.  Finally, linear MR was frequently observed in the past in many
three-dimensional simple metals \cite{parish}. It is now largely accepted that density inhomogeneities are responsible for
this behavior. In a broader sense, this is similar to the present 2D case. However, it is not known whether in 3D metals this
special $R_{xx}$ dependence can be related to $R_{xy}$ in the same way as it seems to be related in our 2DES specimens. Maybe
such an observation will require a special form of inhomogeneity, e.g., a simple density gradient.

In summary, in an ultra high quality two-dimensional electron system, a linear magnetoresistance is observed, which implies
particular ratios for the resistivity of CFs at different, even-denominator filling factors. We show that such a linear $B$
field dependence of $R_{xx}$ cannot be understood within the scattering model for CFs. Rather, it can be reproduced by the
Hall resistance $R_{xy}$, based on an empirical density gradient model.

We thank N. Cooper, H. Fertig, M. Goerbig, W. Kang, E. Rezayi, S. Simon, and A. Stern for helpful discussions. Parts of
measurement were carried out at the National High Magnetic Field Laboratory, which is supported by the NSF, the State of
Florida, and the DOE. We thank G. Jones, T. Murphy, E. Palm, S. Hannahs, and B. Brandt for technical assistance. The work at
Columbia was supported by NSF under No. DMR-03-52738, by DOE under No. DE-AIO2-04ER46133, and by the W.M. Keck Foundation.
The work at Princeton was supported by the AFOSR, the DOE, and the NSF. Sandia is a multiprogram laboratory operated by
Sandia Corporation, a Lockheed-Martin company, for the U.S. Department of Energy.

\end{document}